\documentclass[a4paper,fleqn,usenatbib]{mnras}
\usepackage[T1]{fontenc}
\usepackage{ae,aecompl}

\usepackage{graphicx}	
\usepackage{amsmath}	
\usepackage{amssymb}	

\title[Long-term evolution of RRAT J1819--1458]{Long-term evolution of RRAT J1819--1458}

\author[A. A. Gen\c{c}ali et al.]{
A. A. Gen\c{c}ali,$^{1}$\thanks{E-mail: gencali@sabanciuniv.edu}
\"{U}. Ertan,$^{1}$
\\
$^{1}$Sabanc{\i} University, Orhanl{\i} Tuzla, 34956 \.{I}stanbul, Turkey
}

\date{Accepted XXX. Received YYY; in original form ZZZ}

\pubyear{2018}

\begin{document}
\label{firstpage}
\pagerange{\pageref{firstpage}--\pageref{lastpage}}
\maketitle

\begin{abstract}
At present, J1819--1458 is the only rotating radio transient (RRAT) detected in X-rays. We have studied the long-term evolution of this source in the fallback disc model. The model can reproduce the period, period derivative and X-ray luminosity of J1819--1458 simultaneously in the accretion phase at ages $\sim 2 \times 10^5$~yr. We obtained reasonable model curves with a magnetic dipole field strength $B_0 \sim 5 \times 10^{11}$~G on the pole of the neutron star, which is much weaker than the field inferred from the dipole-torque formula. With this $B_0$ and the measured period, we find J1819--1458 below and close to the radio pulsar death line. Our results are not sensitive to initial period, and the source properties can be produced with a large range of disc masses. Our simulations indicate that J1819--1458 is evolving towards the properties of dim isolated neutron stars at later phases of evolution. This implies a close evolutionary link between RRATs and dim isolated neutron stars. For other RRATs with measured period derivatives and unknown X-ray luminosities, we have estimated the lower limits on the $B_0$ values in the fallback disc model. These limits allow a dipole field distribution for RRATs that could fill the $B_0$ gap between the estimated $B_0$ ranges of dim thermal isolated neutron stars and central compact objects in the same model.
\end{abstract}




\section{Introduction}
\label{intro}

Rotating Radio Transients (RRATs) were discovered more than a decade ago as a new neutron star population \citep{Mc2006}. Unlike normal radio pulsars, RRATs do not exhibit regular radio pulses. They show sporadic and brief radio bursts with time separations of $\sim$ minutes to a few hours. Durations of the radio bursts range from $0.5$ ms to $100$ ms with flux densities from $\sim 10$~mJy to $\sim 10$~Jy, which make these systems the brightest radio sources in the universe \citep{Mc2006, Deneva2009}. Detectable radio emission from a particular RRAT lasts for less than one second per day \citep{Mc2006}. From the analysis of burst times-of-arrival \citep{Manchester2001}, the rotational periods have been obtained in the $0.1 - 7$~s range \citep{Mc2006, Deneva2009}. Among more than 100 confirmed RRATs \citep{Taylor2016}, only J1819--1458 was detected in X-rays \citep{Mc2007}, and upper limits on the X-ray luminosity were estimated for J0847-4316 and J1846-0257 \citep{Kaplan2009}. The main reason for non-detection of the other RRATs in X-rays is the uncertainties in the positions of the sources \citep{Kaplan2009}.

	For J1819--1458 (hereafter J1819), the rotational period $P=4.26$~s \citep{Mc2006} and the period derivative $\dot{P} \approx 5.75 \times 10^{-13}$~s~s$^{-1}$ \citep{Keane2011} give the characteristic age $\tau_{\mathrm{c}} = P/2\dot{P} \simeq 1.2 \times 10^5$~yr and the rotational power $\dot{E} \simeq 4 \pi^2 I \dot{P} P^{-3} \simeq 3 \times 10^{32}$~erg~s$^{-1}$, where $I$ is the moment of inertia of the neutron star. Radio bursts from J1819, repeating about every four minutes, were detected in Parkers observations \citep{Mc2006}. The distance is estimated to be $d = 3.6$~kpc from the dispersion measure with an uncertainty of $\sim 25$\% \citep{Mc2006}. An unabsorbed flux of $1.5 \times 10^{-13}$~erg~s$^{-1}$~cm$^{-2}$ detected in the $0.3-5$~keV band gives an X-ray luminosity $L_{\textmd{x}} = 4 \times 10^{33}~(d/3.6~$kpc$)^2$~erg~s$^{-1}$, which is an order of magnitude higher than the rotational power of the source \citep{Rea2009}. The X-ray spectrum is thermal and can be fitted with a blackbody temperature of $kT \sim 0.14$~keV, $N_\mathrm{H} \sim 6 \times 10^{21}$~cm$^{-2}$ and absorption line at $\sim 1$~keV \citep{Mc2007, Rea2009}. If this is a cyclotron absorption line, the required field strengths are $2 \times 10^{14}$~G and $\sim 10^{11}$~G for the absorption by protons and electrons respectively \citep{Miller2013}.

	The reason for the transient nature of the radio emission from RRATs has not been understood yet. It was proposed that RRATs could have properties similar to the systems that show giant pulses \citep{Knight2006} or to nulling pulsars \citep{Redman2009}. Alternatively, RRATs could be the radio pulsars close to the pulsar death line in the magnetic dipole field-period plane \citep{Chen1993}. In this late phase of radio-pulsar evolution, pulsations might become rare \citep{Zhang2007}. These systems might be emitting weak, continuous radio pulses, which have not been detected yet, in addition to the observed short radio bursts \citep{Weltevrede2006}. It was also proposed that RRATs could have evolutionary links with the anomalous X-ray pulsars (AXPs), soft gamma repeaters (SGRs) \citep{Mc2006, Mc2009} or thermally emitting dim isolated neutron stars (XDINs) \citep{Popov2006}. This possibility has motivated us to study the long-term evolution of J1819 in the fallback disc model that was applied earlier to the other neutron star populations.
	
	 
	 The fallback disc model was first proposed to explain the long-term X-ray luminosity and period evolution of AXPs \citep{Chatterjee2000}. It was proposed by \citet{Alpar2001} that the observed properties of not only AXPs but also other neutron star populations, SGRs, XDINs, and possibly central compact objects (CCOs), could be explained if the fallback disc properties are included in the initial conditions in addition to the magnetic dipole moment and the initial period. To test these ideas, a long-term evolution model for neutron stars with fallback discs was developed including the effects of X-ray irradiation with contribution of the intrinsic cooling of the neutron star, and the inactivation of the disc at low temperatures on the evolution of the star \citep{ErtanE2009, Alpar2011, Caliskan2013}. Later, it was shown that the individual source properties of AXP/SGRs \citep{Benli2016}, XDINs \citep{Ertan2014}, high magnetic-field radio pulsars (HBRPs) \citep{Caliskan2013, Benli2017, Benli2018}, and CCOs \citep{Benli2018_CCOs} can be reproduced in the same long-term evolution model with very similar main disc parameters, supporting the idea proposed by \citet{Alpar2001}. 
	 
	 In this model, estimated magnetic dipole moments of these neutron star populations range from $\sim 10^{29}$~G~cm$^3$ to a few $10^{30}$~G~cm$^3$, which are well below the values inferred from the magnetic dipole torque formula. From the numerical simulations, most AXP/SGRs are estimated to be in the accretion regime, while XDINs are found in the strong propeller regime. In line with these results, it was shown that the characteristic high-energy spectra of AXPs can be produced in the accretion column, consistently with the observed phase dependent pulse profiles \citep{Trumper2010, Trumper2013, Kylafis2014}.
	 
There are several reasons indicating that RRATs could have rather different properties in comparison with the normal radio pulsars, which also motives us to investigate RRATs in the fallback disc model.
If RRATs are the neutron stars evolving in vacuum and spin down with magnetic dipole torques, they would be expected to show regular radio pulses, like many normal radio pulsars with similar rotational properties. They are bright radio emitters but with durations much shorter than their spin periods. While the continuous radio pulses are estimated to cease below the radio pulsar death line, the mechanism producing the radio bursts from RRATs, and when this behavior starts and terminates are not clear yet. These sources could be in an evolutionary phase that starts after the termination of the normal radio pulses. In this situation, they are expected to be close to the pulsar death line (below or above). This could be possible only if they are evolving with the external torques dominating the dipole torques. Because, the dipole fields inferred from the dipole torque formula places them well above the pulsar death line. If these systems are evolving with fallback discs, dipole torque formula could overestimate the actual field by one or two orders of magnitude (see e.g. \citealp{Ertan2014} for XDINs, \citealp{Benli2016} for AXP/SGRs). In this case, these systems could indeed be close to the death line in the $B~-~P$ plane. On the other hand, the thermal X-ray luminosity of J1819 can be emitted only by very young normal radio pulsars with ages less than about $10^4$~y (see Sec. \ref{secmodel}), much smaller than the characteristic age of the source ($> 10^5$~y). Results of our earlier work on the long-term evolution of XDINs show that the normal radio pulsars are not likely to be the progenitors of XDINS, and that there could be evolutionary links between RRATs and XDINs \citep{Ertan2014}. Furthermore, investigations of the statistical, rotational and X-ray properties indicate that RRATs could be progenitors of XDINs \citep{Popov2006}. Investigation of the evolution of J1819 in the fallback disc model could help us understand the evolutionary phase and the conditions that could be responsible for the RRAT behavior, if the source is indeed evolving with a fallback disc. In Section \ref{secmodel}, we briefly describe our model and give the results of the numerical simulations for J1819. We discuss and summarize our conclusions are summarized in Section \ref{secconc}.

\section{The model and application to RRAT J1819--1458}
\label{secmodel}

	Since the details of the model with applications to other neutron star systems are described in the earlier work \citep[see e.g.][]{Ertan2014, Benli2016, Benli2017} here we summarize the initial conditions and the basic disc parameters. To clarify the estimation of the lower limits to the dipole field strengths of RRATs, we also briefly describe the torque calculation employed in the model.

In the fallback disc model, the rotational evolution of the neutron star is governed mainly by the evolution of the disc, irradiated by the X-rays, produced either by mass accretion onto the star or by intrinsic cooling of the star when accretion is not allowed. In the spin-down phase 
there are two basic states: (1) the accretion with spin-down (ASD) state, and (2) the propeller state. In the ASD state, the inner disc interacts with the dipole field of the star in an interaction region (boundary) between the conventional Alfv$\acute{e}$n radius, $r_{\mathrm{A}}$, and the co-rotation radius, $r_{\mathrm{co}}$, at which the field lines co-rotating with the star have the same speed as the Kepler speed of the disc matter. To calculate the magnetic spin-down torque acting on the star we integrate the magnetic torques from $r_{\mathrm{co}}$ to $r_{\mathrm{A}}$ taking $B_\mathrm{z} \simeq B_{\phi}$, where $B_\mathrm{z}$ and $B_{\phi}$ are the poloidal and azimuthal components of the field lines interacting with the inner disc. That is, for the ASD phase, we assume that the inner radius of the boundary region is equal to $r_\mathrm{co}$. The conventional Alfv$\acute{e}$n radius can be written as $r_{\mathrm{A}} \simeq (G M)^{-1/7} \mu^{4/7} \dot{M}_\mathrm{in}^{-2/7}$, where $G$ is the gravitational constant, $M$ and $\mu$ are the mass and the magnetic dipole moment of the neutron star. The integrated magnetic spin-down torque can be written in terms of the disc mass-flow rate, $\dot{M}_\mathrm{in}$, as

\begin{equation}
\label{Nsd}
	N_{\mathrm{SD}} =  \frac{1}{2} ~ \dot{M}_\mathrm{in}  (G M r_\mathrm{A})^{1/2} ~ [1-(r_\mathrm{A}/r_\mathrm{co})^3]  
\end{equation}
\citep{Ertan2008}. When the estimated $r_{\mathrm{A}}$ is greater than the light cylinder radius $r_\mathrm{LC} = c/\Omega_{\ast}$, where $c$ is the speed of light, we replace $r_{\mathrm{A}}$ in equation (\ref{Nsd}) with $r_{\mathrm{LC}}$. In the total torque calculation, we also include the magnetic dipole torque $N_\mathrm{dip} = -2 \mu^2 \Omega_\ast^3 /3c^3$, where $\Omega_\ast$ is the angular frequency of the neutron star, and the spin-up torque resulting from the mass-flow onto the star in the ASD phase, $N_\mathrm{SU} \simeq \dot{M_\ast} (G M r_\mathrm{co})^{1/2}$, where $\dot{M_\ast}$ is the rate of mass accretion onto the star. We calculate the total torque as $N_\mathrm{TOT} = N_\mathrm{SU} + N_\mathrm{dip} + N_\mathrm{SD}$. Over the long-term evolution of AXP/SGRs and XDINs, $N_\mathrm{dip}$ and $N_\mathrm{SU}$ are usually negligible in comparison with $N_\mathrm{SD}$.

	Since the critical condition for transition to the propeller phase is not well known, we use the simplified condition $r_{\mathrm{A}} = r_{\mathrm{LC}}$ for the accretion-propeller transition. Recently, \citet{Ertan2017} estimated the critical accretion rate, $\dot{M}_{\mathrm{crit}}$, for this transition which is consistent with the minimum accretion rates estimated for the transitional millisecond pulsars (tMSPs) \citep[see e.g.][]{Jaodand2016}. The $\dot{M}_{\mathrm{crit}}$ estimated from the observations of tMSPs ($\sim 10^{13}$~g~s$^{-1}$) are much lower than the rates corresponding to $r_{\mathrm{A}} = r_{\mathrm{co}}$, the critical condition for the onset of the propeller phase in the conventional models \citep{Illarionov1975}. Our simplified propeller criterion is roughly in agreement with $\dot{M}_{\mathrm{crit}}$ estimated by \citet{Ertan2017}. In particular, for J1819, our results indicate that the source is currently in the accretion phase with $\dot{M}_{\mathrm{in}} \sim 2 \times 10^{13}$~g~s$^{-1} > \dot{M}_\mathrm{crit} \approx 10^{12}$~g~s$^{-1}$ estimated with the model of \citet{Ertan2017} for $P_0 = 300$~ms and $B_0 \simeq 4.6 \times 10^{11}$~G indicated by our model results (see below). Furthermore, since the onset of the propeller phase corresponds to sharp decay of $L_\mathrm{X}$, the uncertainty in $\dot{M}_{\mathrm{crit}}$ does not affect the model curves significantly.

	Starting from the outermost disc, the disc regions with effective temperature, $T_\mathrm{eff}$, less than a critical temperature $T_\mathrm{P}$ becomes viscously passive. The dynamical outer disc radius $r_{\mathrm{out}}$ is calculated as $r_\mathrm{out} = r (T_\mathrm{eff} = T_\mathrm{P})$. In the long-term evolution, $r_{\mathrm{out}}$ decreases with decreasing X-ray irradiation flux that can be written as $F_{\mathrm{irr}} \simeq 1.2 \: C L_{\mathrm{x}} / (\pi r^2)$ \citep{Fukue1992}, where r is radial distance from the star, $L_{\mathrm{x}}$ is the X-ray luminosity of the star, and $C$ is the irradiation parameter which depends on the disc geometry and the albedo of the disc surfaces. Individual source properties of AXP/SGRs, XDINs, HBRPs, and CCOs could be reproduced self consistently with $T_{\mathrm{P}} \sim 50-150$~K \citep{Benli2016, Benli2017, Benli2018, Benli2018_CCOs} and $C \sim (1-7) \times 10^{-4}$ \citep{Ertan2006, Ertan2007}. The $T_{\mathrm{P}}$ values estimated in our model are in agreement with results indicating that the disc is likely to be active at temperatures $\sim 300$~K \citep{Inutsuka2005}, while our $C$ range is similar to that estimated for the low-mass X-ray binaries \citep[see e.g.][]{Dubus1999}. For the kinematic viscosity, we use the $\alpha$-prescription, $\nu = \alpha \, c_{\mathrm{s}} \, h$ \citep{Shakura1973}, where $c_{\mathrm{s}}$ is the sound speed, $h$ is the pressure scale-height of the disc, and $\alpha$ is the kinematic viscosity parameter.
	
	
\begin{figure}
	\begin{center}
	\includegraphics[width=\columnwidth]{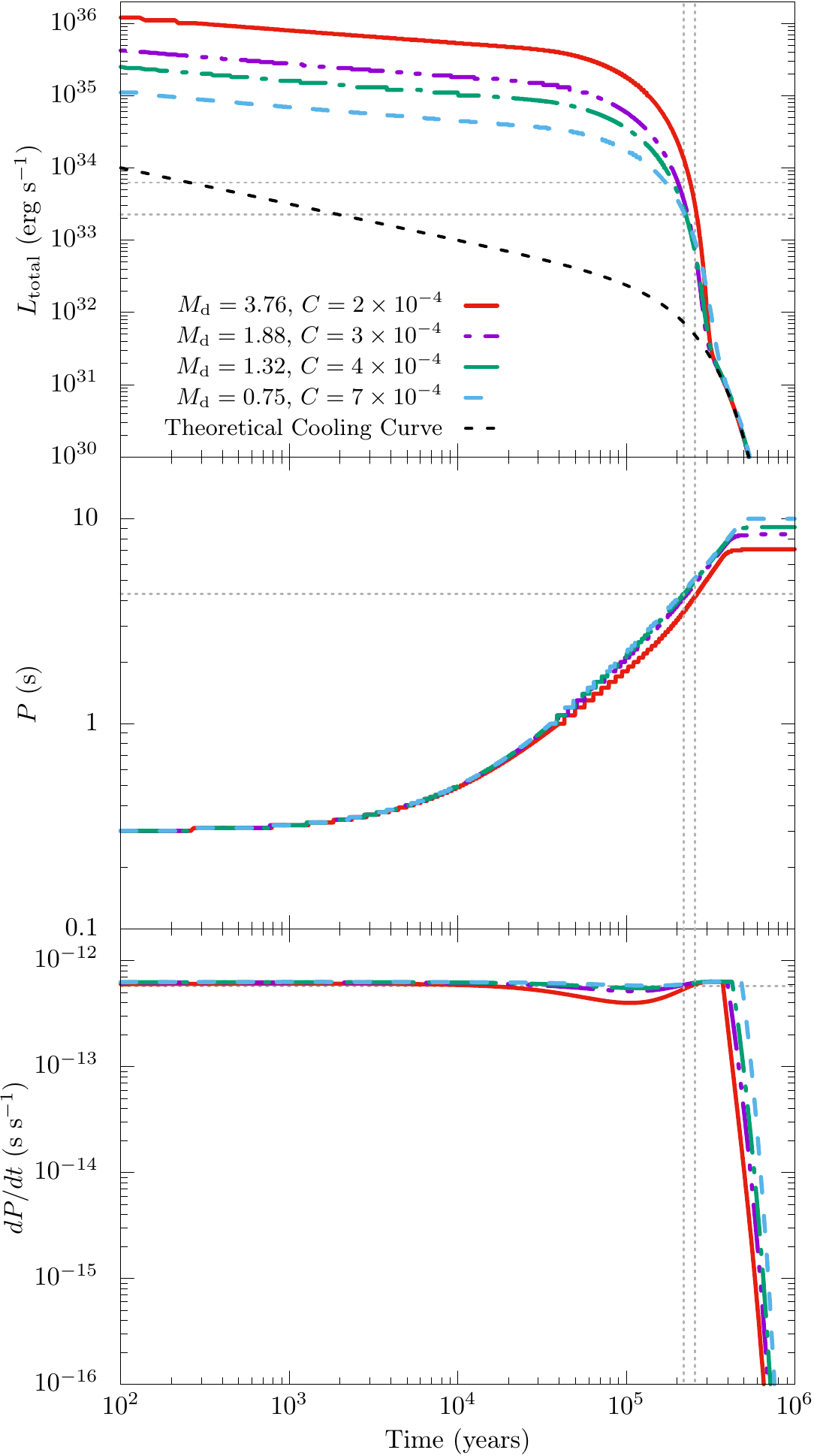}
	\end{center}
    \caption{Illustrative model curves for the long-term evolution of the J1819--1458. These illustrative model curves are obtained with $B_0 = 4.6 \times 10^{11}$~G. The $M_{\mathrm{d}}$ values in units of $10^{-5} M_{\odot}$ and $C$ parameters are given in the top panel. The horizontal dashed lines show the observed $P = 4.26$~s, $\dot{P} \approx 5.75 \times 10^{-13}$~s~s$^{-1}$, and $L_{\mathrm{x}} = 4 \times 10^{33} (d/3.6~$kpc$)^2$~erg~s$^{-1}$ with $25$~\% uncertainty \citep{Mc2006,Keane2011, Rea2009}. For all these curves, $\alpha = 0.045$ and $T_{\mathrm{P}} = 53$~K (see the text for details). Dotted curve in the top panel represents the evolution of the cooling luminosity for a neutron star with a conventional dipole field \citep{Page2009}.}
    \label{fig:J1819_all}
\end{figure}


	The main disc parameters, $\alpha$, $C$, and $T_{\mathrm{P}}$ are similar for the fallback discs in different neutron star populations. The initial conditions, namely the strength of the magnetic dipole field on the pole of the star, $B_0$, the initial rotational period, $P_0$, and the initial mass of the disc, $M_{\mathrm{d}}$, are mainly responsible for rather different characteristics  emerging during the evolutionary phases of the sources. Through many simulations, we determine the allowed ranges of the initial conditions that can produce the $P$, $\dot{P}$, and $L_{\mathrm{x}}$ of sources simultaneously. In most cases, the long-term evolution is not sensitive to $P_0$ \citep[see][for details]{ErtanE2009}. In the present case, we take $P_0 = 300$~ms, the center of the Gaussian distribution estimated for the initial periods of the radio pulsars \citep{Faucher2006}. In Fig. \ref{fig:J1819_all}, we give illustrative model curves that can represent the long-term evolution of J1819. We obtain these model curves with $T_{\mathrm{P}} = 53$~K, $C = (2-7) \times 10^{-4}$, and $\alpha = 0.045$, which are the typical values used in all earlier applications of the same model to AXP/SGR, XDINs and HBRPs \citep[see e.g.][]{Benli2016}. The illustrative sources in Fig. \ref{fig:J1819_all} reach the observed $P$, $\dot{P}$, and $L_{\mathrm{X}}$ of J1819 at an age of $\sim 2 \times 10^5$~yr, when the source is evolving in the accretion phase. The inner radius of the disc is more than $2$ orders of magnitude greater than the radius of the star. That is, in the accretion phase the main source of the X-rays is the accretion onto the neutron star, while the contribution of the inner disc to the X-ray luminosity is negligible. The model constrains $B_0$ to a rather narrow range around $\sim 5 \times 10^{11}$~G, while the source properties can be reproduced with a large range of disc masses, $M_\mathrm{d}$ (see Fig. \ref{fig:J1819_all}). 

\begin{figure}
	\includegraphics[width=\columnwidth]{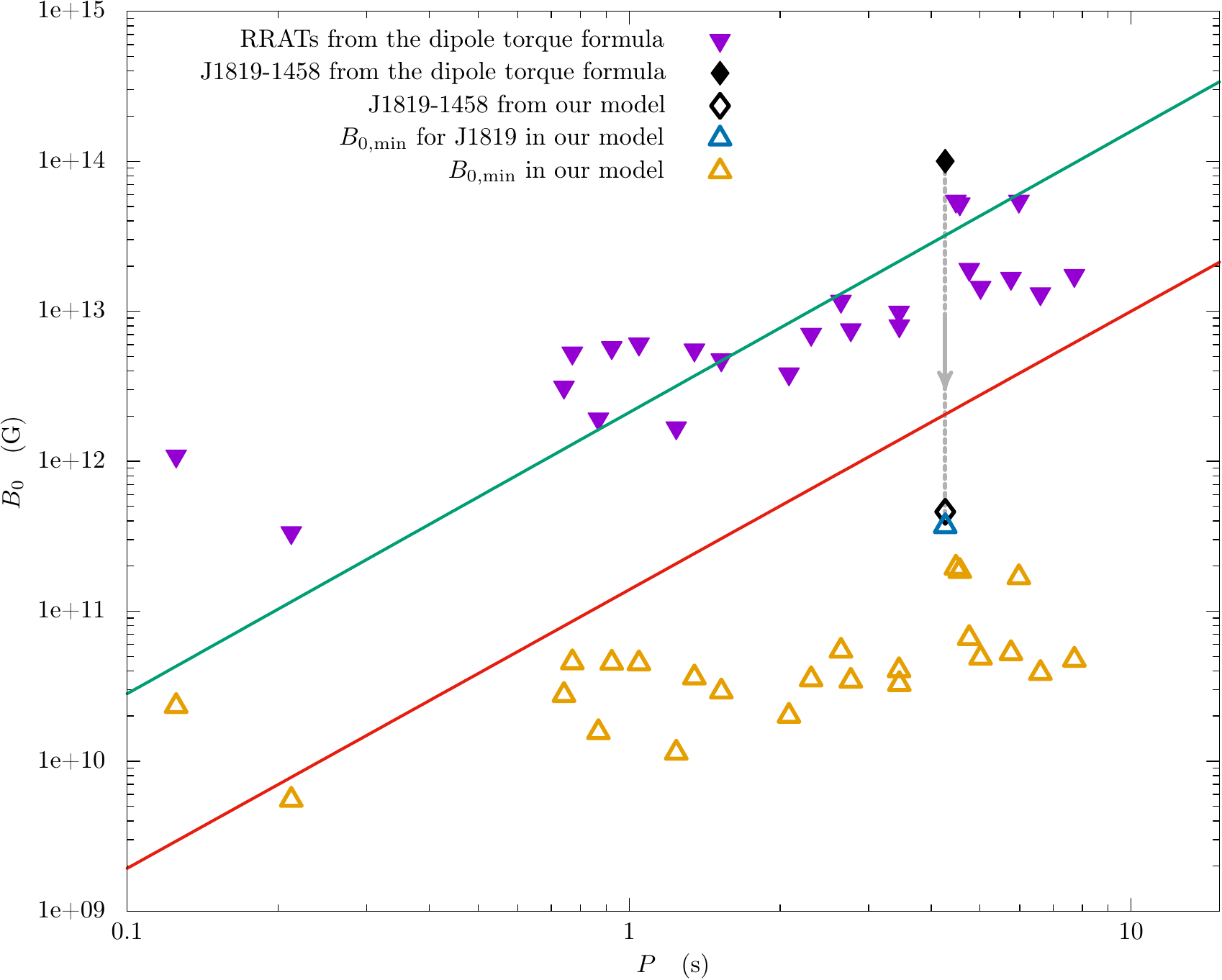}
    \caption{$B_0$ - $P$ diagram. Filled and open diamonds show the $B_0$ values for J1819 inferred from the dipole torque formula ($\simeq 10^{14}$~G) and estimated in our model ($\simeq 4.6 \times 10^{11}$~G) respectively. The minimum $B_0$ ($B_{0,\mathrm{min}}$) values estimated in our model for the other RRATs with known $P$ and $\dot{P}$ \citep{Mc2006, Deneva2009, Burke2010, Keane2010, Keane2011, Burke2011} are marked with open triangles using $B_{0,\mathrm{min}} \simeq 1.5~\dot{P}_{-11}^{1/2}~10^{12}$~G (see the text). For each of these sources, $B_0$ inferred from the dipole torque formula are also plotted (inverse filled triangle). Solid lines represent the borders of the death valley \citep{Chen1993}. The lower border is similar to the classical pulsar death line \citep{Bhattacharya1992}.    
    }
    \label{fig:B0_P}
\end{figure}

	What is the basic, common property causing RRATs to produce radio bursts, and no regular radio pulsations? The dipole field strength indicated by the model results and the measured period place J1819 below and close to the pulsar death line in the $B_0 - P$ plane (Fig. \ref{fig:B0_P}). The model source is evolving into the properties of XDINs, which do not show RRAT behavior. It is not clear whether all RRATs are close to and below the pulsar death line. For RRATs other than J1819, for which the X-ray luminosity is not detected, it is not easy to pin down the evolutionary status with $P$ and $\dot{P}$ alone. Nevertheless, we can estimate the lower bounds on $B_0$ ($B_{0,\mathrm{min}}$), for the sources with measured $\dot{P}$. In our model, the maximum spin-down torque is obtained in the accretion with spin-down (ASD) phase when the source is not very close to rotational equilibrium. This corresponds to the constant $\dot{P}$ phase (see Fig. \ref{fig:J1819_all}) over which the second (negative) term of the magnetic spin-down torque (equation \ref{Nsd}) dominates both the accretion torque and the magnetic dipole torque. For this phase of evolution, it can be seen from equation (\ref{Nsd}) that the torque is independent of $\dot{M}_{\mathrm{in}}$, and the minimum dipole field strength on the pole of the star can be estimated as $B_0(B_{0,\mathrm{min}}) \simeq 1.5 ~ \dot{P}^{1/2}_{-11}~10^{12}$~G where $\dot{P}^{1/2}_{-11}$ is the period derivative in $10^{-11}$~s~s$^{-1}$. This formula, which underestimates the $B_0$ values for the sources that are in the propeller phase or close to the rotational equilibrium in the accretion phase, gives the minimum possible field strength for a given $\dot{P}$ independently of $\dot{M}_{\mathrm{in}}$. These $B_{0,\mathrm{min}}$ values are plotted in Fig. \ref{fig:B0_P}.
	
	Without X-ray luminosity information, we cannot estimate the actual field strength $B_0$. If the RRAT behavior of the sources start when they are close to the pulsar death line, the actual $B_0$ is likely to be between $B_{0,\mathrm{min}}$ and the $B_0$ corresponding to the period of the source on the pulsar death line. The estimated $B_{0,\mathrm{min}}$ values seen in Fig. \ref{fig:B0_P} is important in that it is compatible with a continuous distribution for the $B_0$ values of all single neutron star populations (AXP/SGR, XDIN, HBRP, RRAT and CCO) in the fallback disc model, filling the gap between $B_0~\sim 10^9$~G for CCOs \citep{Benli2018_CCOs} and $B_0~\gtrsim~10^{11}$~G for the other populations \citep{Alpar2001, Eksi2003, Ertan2007, ErtanE2009, Ertan2014, Ertan2017, Benli2017, Benli2018}.
	

\section{Discussions and Conclusions}
\label{secconc}

	We have investigated the long-term evolution of J1819--1458 which is the only RRAT detected in X-rays. We have shown that the period, period derivative and X-ray luminosity of the source can be explained in the same model that can account for the long-term evolutions of AXP/SGRs, XDINs, HBRPs, and CCOs. The model can reproduce the properties of the source only with a narrow range of $B_0$ around $4.6 \times 10^{11}$~G, while reasonable model curves are obtained with rather different initial disc masses (($0.75 - 3.76) \times 10^{-5}~M_\odot$). The model sources reach the properties of J1819 in the accretion with spin-down (ASD) phase at an age $\sim 2 \times 10^5$~yr, when the estimated cooling luminosity of the neutron stars is a few per cent of the observed $L_\mathrm{X}$ of J1819. In the accretion phase, the mass-flow onto the neutron star is expected to switch off the radio pulses. Even if the accretion stops by some reason, we do not expect regular pulsed radio emission from J1819, since the $B_0$ indicated by our model and the measured $P$ place the source below the pulsar death line. 
	
	If the absorption feature around $1$~keV is a proton cyclotron line, the required field strength is $\sim 2 \times 10^{14}$~G \citep{Miller2013}. The field close to the surface of the star could indeed be much stronger than the dipole component due to presence of local strong quadrupole fields. Alternatively, the observed absorption feature could be electron cyclotron line which could be produced in the accretion column with a field strength $ \sim 10^{11}$~G.
	
	Illustrative model curves in Fig. \ref{fig:J1819_all} imply that J1819 is currently evolving through lower part of the AXP/SGR region in the $P$~--~$\dot{P}$ diagram. Currently, the short-term timing behavior of the source seems to have been affected by the glitch effects \citep{Bhattacharyya2018}. From the model results, we estimate that J1819 will reach the XDIN properties within a few $10^5$~yr (Fig. \ref{fig:J1819_all}).

	The illustrative model curves in Fig. \ref{fig:J1819_all} imply that the source is evolving into the XDIN properties. This result is not very sensitive to the initial period, the disc mass and the resultant $L_\mathrm{X}$ history of the source. For the other RRATs, since the X-ray luminosities are not known, it is not possible to estimate their evolutionary paths and the $B_0$ values. Nevertheless, the lower bound, $B_{0,\mathrm{min}}$ for a given source can be estimated using the most efficient torque reached in the ASD phase and the measured $\dot{P}$ of the source (Section \ref{secconc}). In Fig. \ref{fig:B0_P}, it is seen that these lower limits on $B_0$ allow a continuous $B_0$ distribution from CCOs to AXP/SGRs in the fallback disc model.
	
An X-ray nebula was detected around J1819--1458 \citep{Rea2009}. The size of the X-ray nebula is about $1$~ly \citep{Camero2013} which is too large to be related to the outer regions of a fallback disc that could scatter a fraction of the X-rays emitted by the star. The estimated luminosity of the point source is more than 2 orders of magnitude higher than that of the nebula \citep{Rea2009}. The extended emission might be powered by a fraction of the luminosity of the star, nevertheless the mechanism producing the extended emission is not clear yet. In any case, this extended emission with a low luminosity and a very large size could only be a secondary process that does not affect our results obtained for the point source and its interaction with the inner disc with a radius of less than about $10^9$~cm.

It is expected that the pulsed radio emission is quenched by mass-flow on to the star. The behavior of RRATs could also be indicating a mechanism that tries to impede continuous pulsed radio emission, causing the observed sporadic radio emission.  Alternatively, it could be the case that some mechanism could be inducing emission of these radio bursts in a sporadic way. The estimated evolution of J1819 toward the XDIN population might indicate that all known XDINs could have evolved through RRAT phase in the past. The fact that all measured RRAT periods are smaller than $7$ s, and that $5$ out of $7$ XDINs have periods greater than $7$ s could point to a maximum period (for a given $B_0$) above which RRAT behavior disappear. Considering that we have found J1819 below the death line, for a given source, there could be a certain RRAT phase that starts after the termination of the normal radio pulsations, and ends above a critical $P$ for this particular neutron star. It is not clear whether the RRAT behavior itself is related to presence or properties of fallback disc around the source. We need further detections of RRATs in X-rays to test these ideas in depth through long-term evolutionary analysis of these sources.

Unlike J1819, most of the RRATs have characteristic ages greater than a few $10^6$~y. For these sources, if the actual ages are close to the characteristic ages, the cooling luminosities are estimated to be too low to be detected in X-rays.  In our model, these RRATs (except very young systems) are likely to be evolving in the propeller phase at ages much smaller than their characteristic ages, similar to the case estimated for XDINs \citep{Ertan2014}. This means that these systems could have cooling luminosities much greater than those estimated for their characteristic ages, if they are indeed evolving with fallback discs. This prediction of the model can be tested by future detections of RRATs in X-rays.

\section*{Acknowledgements}

We acknowledge research support from Sabanc{\i} University, and from T\"{U}B\.{I}TAK (The Scientific and Technological Research Council of Turkey) through grant 117F144. We thank M. Ali Alpar for useful comments on the manuscript.




\bibliographystyle{mnras}
\bibliography{mnras} 




\bsp	
\label{lastpage}
\end{document}